\documentclass[sigconf]{acmart}
\usepackage{geometry}
\usepackage[medium,compact]{titlesec}
\usepackage{marvosym}
\usepackage{tabularx}
\usepackage{multirow}
\usepackage{multirow}
\usepackage{graphicx}
\usepackage{hyperref}

\acmYear{2024} 
\setcopyright{rightsretained} 
\acmConference[ICCAD '24]{IEEE/ACM International Conference on
 Computer-Aided Design}{October 27--31, 2024}{New York, NY, USA}
 \acmBooktitle{IEEE/ACM International Conference on Computer-Aided Design
 (ICCAD '24), October 27--31, 2024, New York, NY, USA}
 \acmDOI{10.1145/3676536.3676679}
 \acmISBN{979-8-4007-1077-3/24/10}
\begin{document}

\title[]{Natural language is not enough: Benchmarking multi-modal generative AI for Verilog generation}

\begin{abstract}
Natural language interfaces have exhibited considerable potential in the automation of Verilog generation derived from high-level specifications through the utilization of large language models, garnering significant attention. Nevertheless, this paper elucidates that visual representations contribute essential contextual information critical to design intent for hardware architectures possessing spatial complexity, potentially surpassing the efficacy of natural-language-only inputs. Expanding upon this premise, our paper introduces an open-source benchmark\footnote{\textbf{\url{https://github.com/aichipdesign/chipgptv}}} for multi-modal generative models tailored for Verilog synthesis from visual-linguistic inputs, addressing both singular and complex modules. 
Additionally, we introduce an open-source visual and natural language Verilog query language framework to facilitate efficient and user-friendly multi-modal queries.
To evaluate the performance of the proposed multi-modal hardware generative AI in Verilog generation tasks, we compare it with a popular method that relies solely on natural language. Our results demonstrate a significant accuracy improvement in the multi-modal generated Verilog compared to queries based solely on natural language. We hope to reveal a new approach to hardware design in the large-hardware-design-model era, thereby fostering a more diversified and productive approach to hardware design.
\end{abstract}

\author{\normalsize
\textbf{\textbf{Kaiyan Chang}\textsuperscript{1,3}, Zhirong Chen\textsuperscript{1,2,3},  
Yunhao Zhou\textsuperscript{2,5},Wenlong Zhu\textsuperscript{1,2,3}, Kun Wang\textsuperscript{1,2,4}, Haobo Xu\textsuperscript{1,2,3}, Cangyuan Li\textsuperscript{1,2,3}, Mengdi Wang\textsuperscript{1,2},
Shengwen Liang\textsuperscript{1},  Huawei Li\textsuperscript{1,3},Yinhe Han\textsuperscript{1,2},{Ying Wang*}\Letter\textsuperscript{1,2}}\\
\textit{SKLP, Institute of Computing Technology, Chinese Academy of Sciences, Beijing, China\textsuperscript{1}\\
CICS, Institute of Computing Technology, Chinese Academy of Sciences, Beijing, China\textsuperscript{2}\\
University of Chinese Academy of Sciences\textsuperscript{3} \\
Hangzhou Institute for Advanced Study, University of Chinese Academy of Sciences\textsuperscript{4}\\
Shanghai Jiao Tong University\textsuperscript{5}\\
changkaiyan@live.com, wangying2009@ict.ac.cn
}}
\authornote{Ying Wang is the corresponding author.}

\maketitle

\section{Introduction}

Recent advancements in the deployment of large language models for Verilog generation have garnered significant attention within the realm of electronic design automation (EDA) \cite{benchmarking,chipnemo,chen2024dawn,chang2023chipgpt}. These models are emerging as a pivotal methodology for the automated generation of both Verilog code and EDA scripts \cite{verigen,blocklove2023chip,chipnemo}, which heralds a transformative shift in the area. The principal objective of this line of inquiry is to enable hardware developers to quickly design intricate hardware systems without requiring extensive expertise in the specific hardware \cite{fu2023gpt4aigchip}.
%Recent advancements in utilizing large language models for Verilog generation have sparked significant interest within the field of electronic design automation。
%without requiring extensive knowledge of the specific hardware

%无法解决的场景：状态机、多模块生成问题。核心问题：自然语言瓶颈问题。最终方案：使用多模态大模型。

%现在多模态大模型的挑战：没有benchmark评价哪个多模态大模型比较好, 缺少一个快速benchmarking操作模型和数据的方案，缺少对benchmark的定义。
Although natural language interfaces have shown potential to perform fundamental code generation tasks, the domain of hardware design presents distinct and intricate challenges that transcend the capabilities of linear linguistic representations. Significant obstacles exist in the generation of intricate architectures involving state machines and the integration of multiple interacting modules, as illustrated in Fig. \ref{fig:multimodulecase} and Fig. \ref{fig:casestudyresult}. One critical limitation is the challenge of effectively conveying the spatial interrelations and complex nested configurations inherent in hardware designs, attributed to the inherently sequential nature of language. Our empirical studies suggest that a multi-modal generative model, incorporating linguistic elements with visual block diagrams and structural data, holds promise in addressing these constraints.

While our experiments substantiate the potential of multi-modal generative models in the domain of hardware design, three pivotal challenges remain to be addressed. A fundamental issue is the absence of standardized benchmarks, essential to assess and contrast various multi-modal architectures. In the absence of universally accepted benchmarks and metrics, it is arduous to perform systematic evaluations of model performance or to foster advancements in the field. Moreover, the current practice of benchmarking multi-modal models necessitates extensive labeled datasets and a protracted training process, hampering rapid iteration and the exploration of innovative model paradigms. There is a critical need for benchmarking frameworks that facilitate model evaluations with smaller data sets or restricted supervision. Lastly, the concept of multi-modality in the context of hardware tasks requires a precise definition. Important questions remain about which modalities are the most effective, how they should be integrated and processed, and how to quantitatively evaluate their performance on complex generative tasks. In response to these challenges, we introduce a Verilog multi-modal benchmark, a Verilog multi-modal model query language, and a comprehensive specification of the benchmark.

\begin{table}[ht]
\caption{Related Work Comparasion. Natural Language and Image Co-generation is a zero-shot method with a higher level.}
\label{tab:relatedwork}
\resizebox{\linewidth}{!}{%
\begin{tabular}{|l|l|l|l|l|}
\hline
\textbf{\textbf{Works}} &
  \textbf{Task} &
  \textbf{Input} &
  \textbf{Output} &
  \textbf{Method} \\ \hline
Vivado &
  \begin{tabular}[c]{@{}l@{}}Block \\ Design\end{tabular} &
  \begin{tabular}[c]{@{}l@{}}Diagram \\ Template\end{tabular} &
  Verilog &
  Rule-based \\ \hline
Qaw\cite{notationquantum} &
  \begin{tabular}[c]{@{}l@{}}Notation \\ Programming\end{tabular} &
  \begin{tabular}[c]{@{}l@{}}Quantum \\ Notation\end{tabular} &
  QASM &
  \begin{tabular}[c]{@{}l@{}}Object \\ Detecting\end{tabular} \\ \hline
ChipNeMo\cite{chipnemo} &
  \begin{tabular}[c]{@{}l@{}}Verilog \\ Generation\end{tabular} &
  \begin{tabular}[c]{@{}l@{}}Natural \\ Language\end{tabular} &
  Verilog &
  LLM \\ \hline
Thakur et al.\cite{benchmarking} &
  \begin{tabular}[c]{@{}l@{}}Verilog \\ Generation\end{tabular} &
  \begin{tabular}[c]{@{}l@{}}Natural \\ Language\end{tabular} &
  Verilog &
  LLM \\ \hline
Ours &
  \begin{tabular}[c]{@{}l@{}}Verilog \\ Generation\end{tabular} &
  \begin{tabular}[c]{@{}l@{}}Img. + \\ Natural Lang.\end{tabular} &
  Verilog &
  multi-modal \\ \hline
\end{tabular}%
}
\end{table}
\vspace{-1mm}
In this paper, we elucidate the limitations inherent in relying exclusively on natural language-based Verilog generation methodologies and address the aforementioned challenges. We introduce a novel co-generation query language to facilitate benchmarking, meticulously designed to formalize the automatic co-generation process intrinsic to our proposed method. Moreover, to aid in the comparison and evaluation of prospective works in this nascent field, we propose a benchmark with precise specifications for future investigations in the domain of vision-language Verilog models. The evaluation is conducted using GPT4V, GPT4, LLaVA, and LLaMA, which are representative foundational models for natural language Verilog generation\cite{blocklove2023chip}. The results demonstrate that multi-modal large language models for Verilog generation exhibit substantial improvements over conventional methodologies. In summary, we aspire to unveil a novel field in the era of large hardware design models, thereby promoting a more diversified and efficacious approach to hardware design.
% 现在大模型还没有能完整复现出fsm的能力 只能做到大致正确
The contributions are listed below:
\begin{itemize}
    %reveal，finding结论
    %\item We introduce vision representation to the Verilog generation by large language model, and show that visual representations provide additional context critical for design intent.
    \item We disclose that for hardware structures with spatial complexity, the visual representations presented in this paper offer crucial additional context to clarify design intent, potentially surpassing natural-language-only inputs in AI-driven automated hardware generation.
    \item \item We introduce an innovative multimodal model query language designed to formalize vision-language descriptions, effectively reducing token cost while enhancing the quality of the generated code.
    \item We present a hierarchical benchmark, ranging from simple to complex, to assess the performance of multimodal large models in Verilog generation. This benchmark will be made open-source in conjunction with the proposed multimodal query language framework.
    \item We conduct a systematic evaluation of multimodal large models using our benchmark, encompassing syntax, functionality, and next-token success rate. Incorporation of visual representations significantly improves the testbench passing rate from 46.88\% to 71.81\% for the GPT4 series, and from 13.41\% to 25.88\% for the LLaMA series.

% 关于规范的贡献，以及规范带来的效果提升？
\end{itemize}

\section{Background \& Motivation}
\label{sec:moti}
\subsection{The limitations in current LLM-based RTL generation roadmap.}

Researchers have examined the utilization of large language models (LLMs) in the context of Verilog code generation, as depicted in Tab. \ref{tab:relatedwork}. Benchmarking results presented in \cite{benchmarking,lu2023rtllm,liu2023verilogeval} elucidate the potential of these models to mitigate the challenges faced by hardware designers. Significant advancements have been achieved in the domain of fine-tuning for code completion \cite{verigen,chang2024chipgptft}, general RTL generation \cite{blocklove2023chip}, and the generation of EDA tool scripts \cite{chipnemo}. Beyond single-sentence models such as GPT-3, conversational large language models (LLMs) have demonstrated proficiency in a variety of advanced applications, including RTL-level repair \cite{blocklove2023chip,tsai2023rtlfixer,fang2024assertllm,yao2024hdldebugger}, quantum computing \cite{quantumllm}, in-memory computing \cite{cimllm}, testing \cite{fixing,llmassertion,orenesvera2023rtl}, and the design of AI domain-specific processors \cite{fu2023gpt4aigchip}. Nonetheless, the generation of RTL utilizing LLMs remains constrained by several limitations:

\begin{figure}
    \centering
    \includegraphics[width=\linewidth]{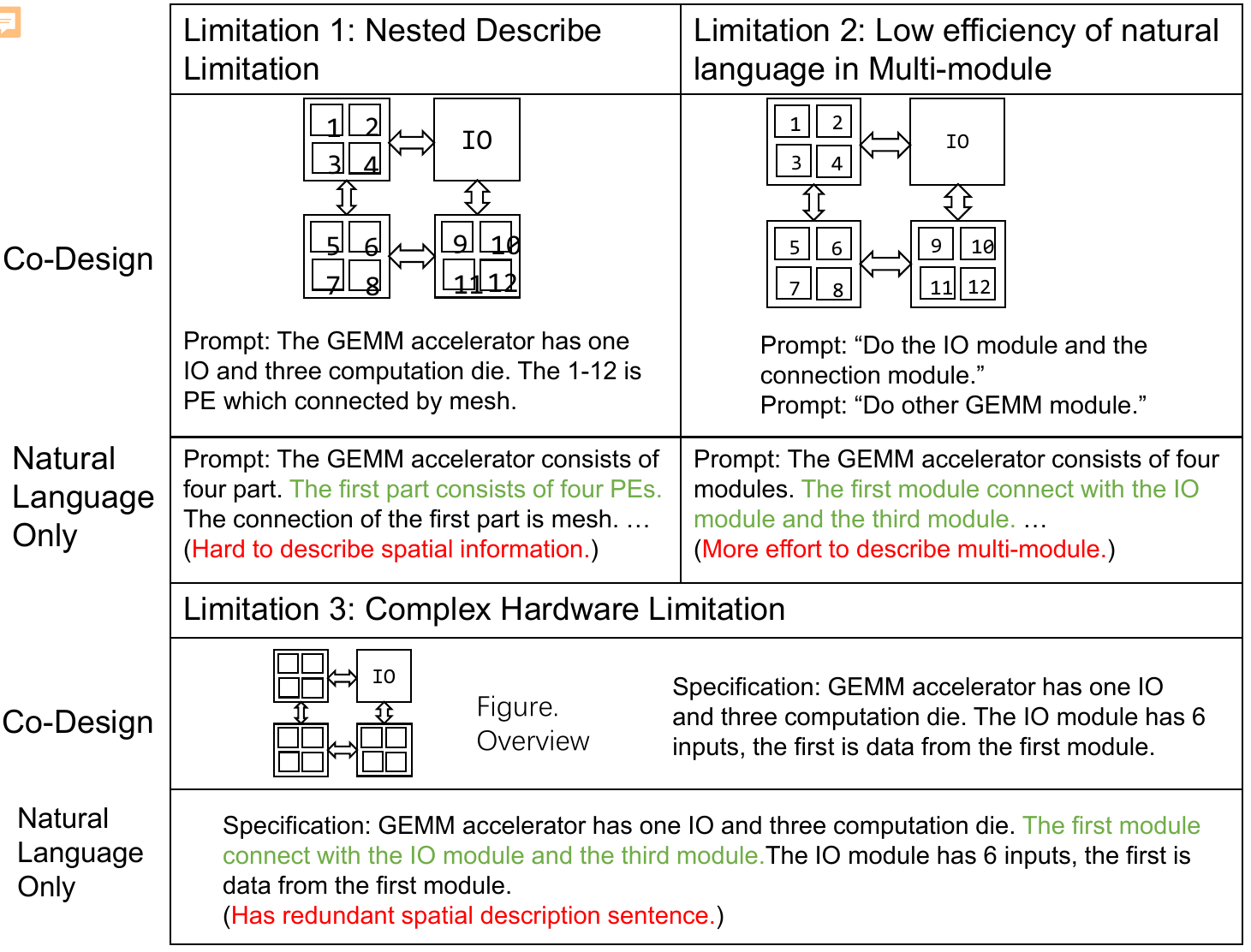}
    \caption{Constraints in Hardware Design Using Natural Language: An Analysis from Three Perspectives. The co-design row illustrates solutions leveraging the multi-modal approach. The "natural language only" row represents results derived from a text-only language model. The green sentence indicates redundant information that can be more efficiently conveyed through visual representation. }
    \label{fig:limitation}
\end{figure}
\paragraph{Limitation 1: The inherent limitations of linguistic representations render natural language insufficient for accurately conveying the nested spatial structures of hardware.} The intricate spatial relationships among components in computer hardware cannot be fully articulated using solely natural language descriptions. For instance, the spatial accelerator depicted in Fig. \ref{fig:limitation} encompasses the IO module, PE, and controllers. The spatial interconnections and multi-tiered component structures rapidly exceed the representational capacity of natural language. Terms such as "below," "embedded," "adjacent," and "surrounding" fall short in adequately expressing the embedded 2D spatial structures. Although natural language can depict basic hardware organization, its linear and imprecise representations lack the relational expressiveness requisite for modeling multi-level spatial information. Hence, formal diagrams are indispensable for comprehensively representing structural complexity.

\paragraph{Limitation 2: Inefficiency of Linguistic Descriptions in Multi-Module Hardware Design} Within the realm of advanced hardware design, particularly when analyzing systems composed of up to $n$ submodules, traditional linguistic methodologies for articulating interconnections demonstrate notable inefficiencies, as illustrated for $n=12$ in Fig. \ref{fig:limitation}. Such hardware configurations can be represented as a multigraph $G = (V, E)$, where each submodule is depicted as a node and the interconnections among these submodules are denoted as edges. The complexity of this system can escalate to the order of $O(n^2)$ in terms of interconnections, thus necessitating a description complexity of $O(n^2)$ tokens. This condition highlights a fundamental limitation inherent to single-modal language models, which are markedly less efficient than multi-modal models when managing the intricate details of interconnectivity within hardware designs.
\paragraph{Limitation 3: Risk of Misalignment in Complex Hardware Designs Using Language Models} Misalignment presents a significant obstacle in the deployment of large language models for intricate hardware design. This complication typically arises when the descriptions provided are ambiguous or incomplete. A particularly salient example of this issue is the port connection, as depicted in Fig. \ref{fig:limitation}. Sole reliance on textual input to specify the functional elements of a design often leads to the model misrouting signals to incorrect ports or unintended submodules, thereby deviating from the expected configuration. This misalignment risk is substantially ameliorated through the incorporation of visual inputs. By adopting a multi-modal approach that encompasses visual information, the model obtains explicit visual cues that facilitate the correct alignment of signals to their designated ports. This enhancement not only clarifies the intended design but also markedly diminishes the probability of alignment errors, thus enhancing the accuracy and dependability of the model's output in hardware design applications.

\subsection{Visual and Natural Language Hardware Co-Design Case Study} 
\begin{figure}[htbp]
    \centering
    \includegraphics[width=\linewidth]{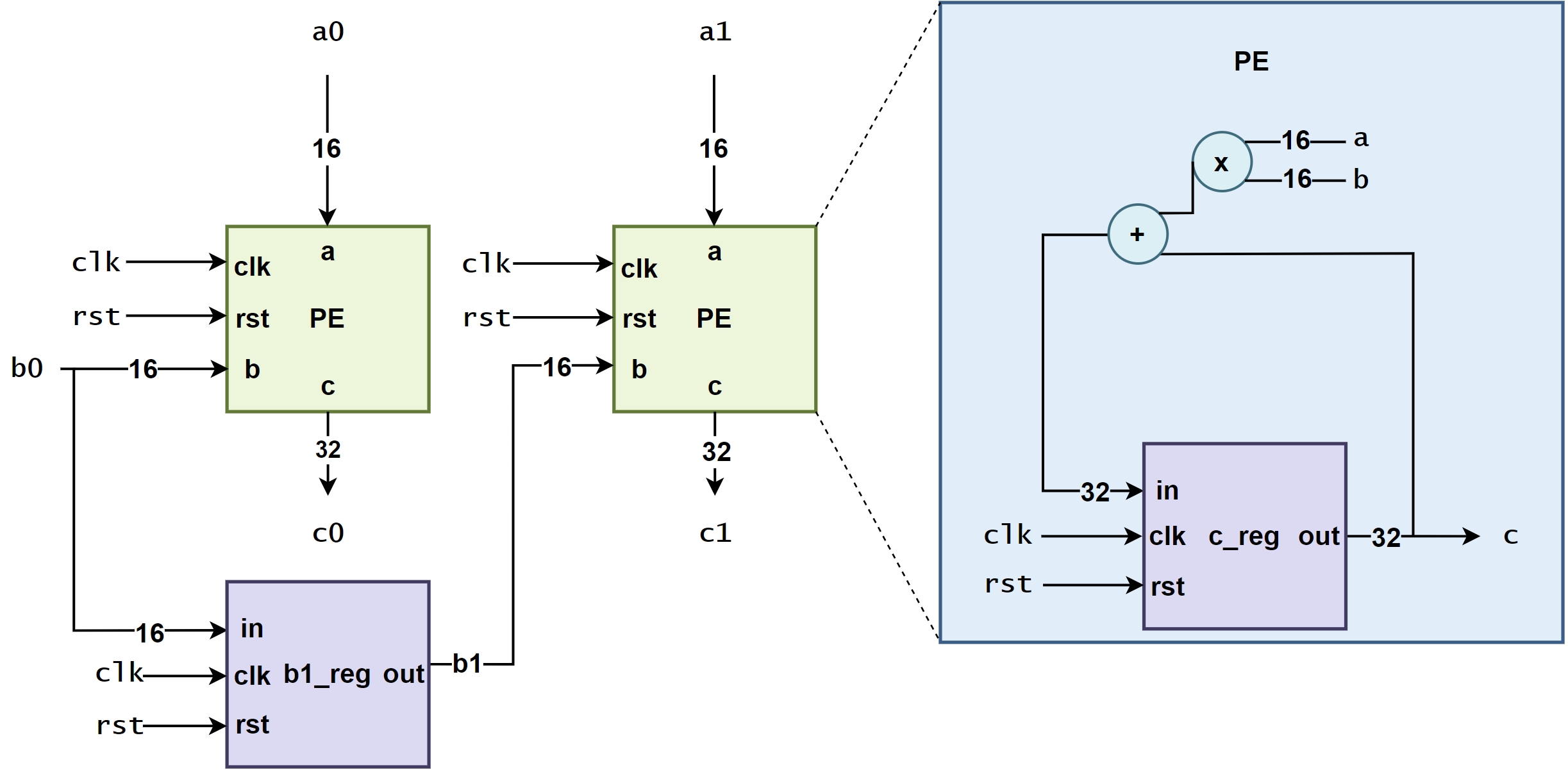}
    \caption{A case study to show a multiply and accumulate PE.}
    \label{fig:nocpestructure}
\end{figure}

\begin{figure}
    \centering
    \includegraphics[width=\linewidth]{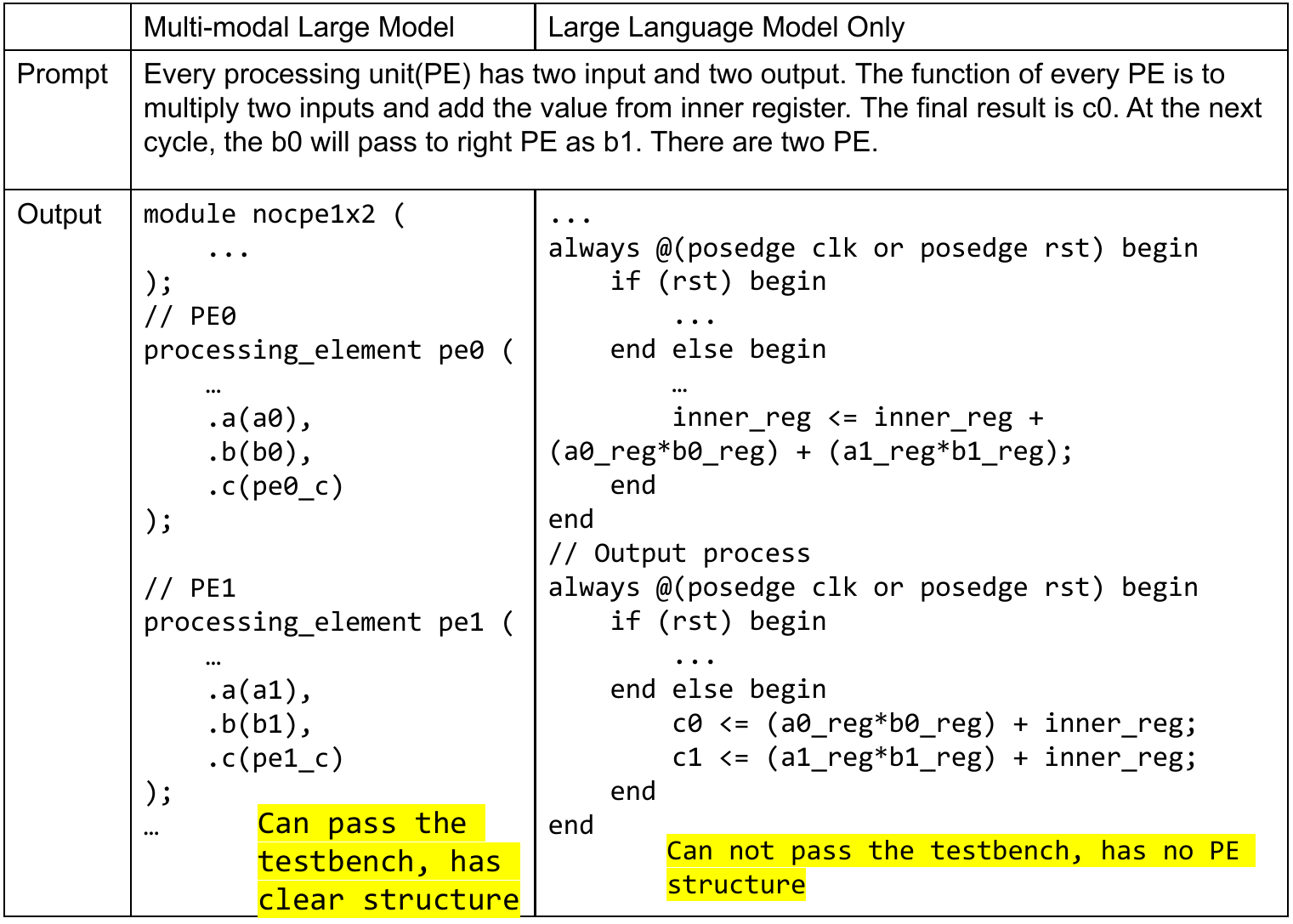}
    \caption{A comprehensive case study illustrating that, within the context of multi-module hardware, the multi-modal model exhibits a substantial enhancement in performance compared to the conventional language model.}
    \label{fig:multimodulecase}
\end{figure}

\begin{figure}
    \centering
    \includegraphics[width=\linewidth]{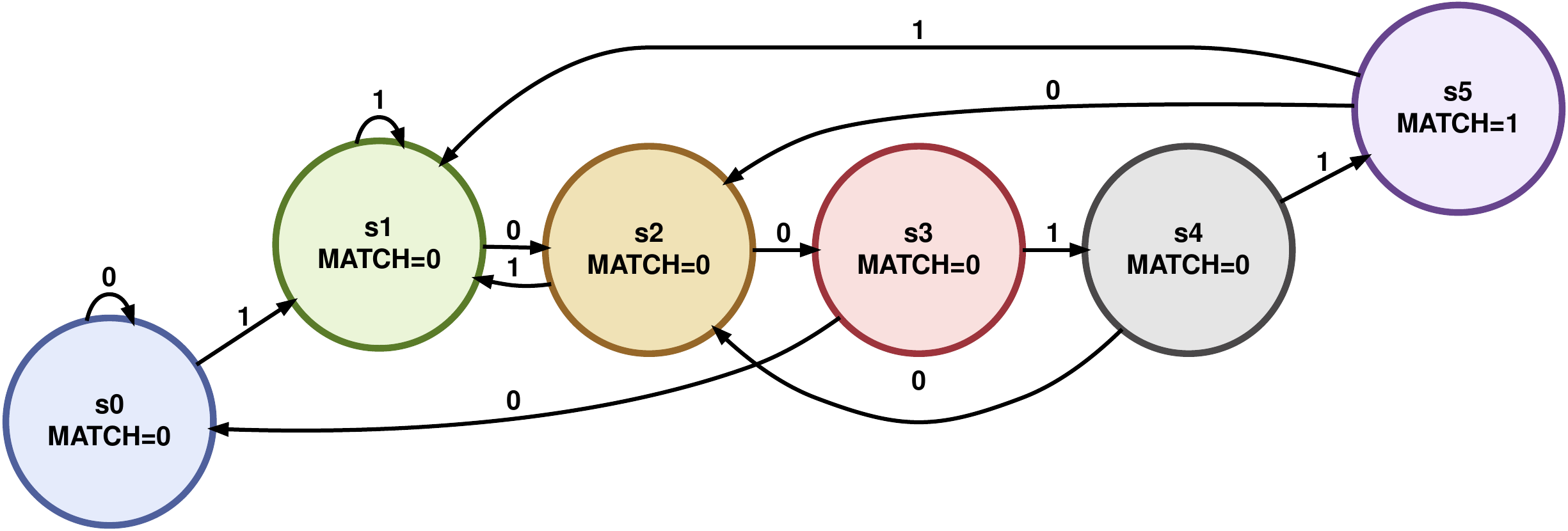}
    \caption{A comprehensive analysis of a state machine elucidating the superiority of integrated visual and natural language co-design over conventional language models. This figure illustrates a state machine designed to identify the input sequence 10011. }
    \label{fig:statemachineexample}
\end{figure}

\begin{figure}
    \centering
    \includegraphics[width=\linewidth]{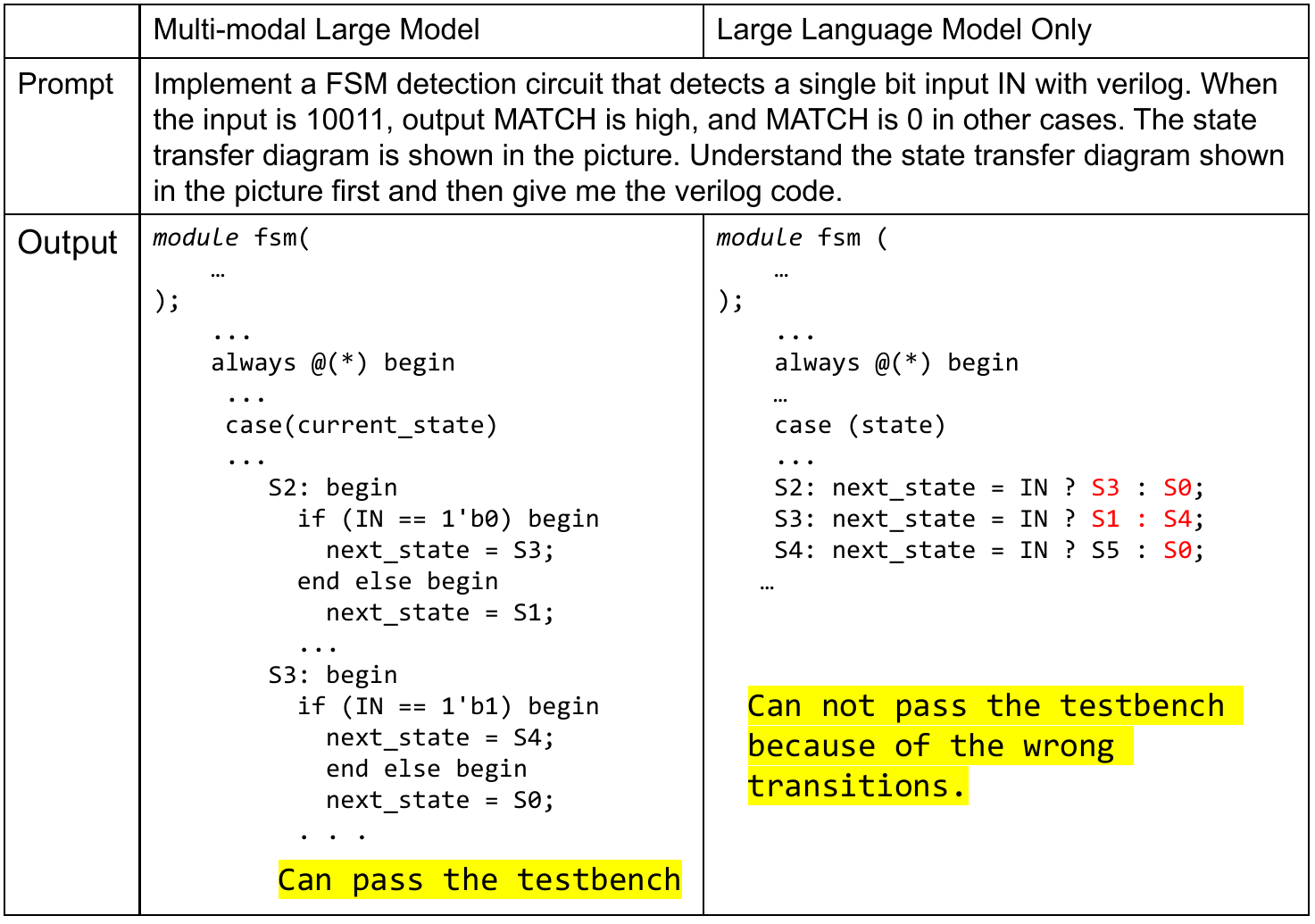}
    \caption{Compare the capability of generating Verilog between multi-modal and natural-language-only model.  To follow a fair standard, we generate a Verilog description and test the generated code using pass@5.}
    \label{fig:casestudyresult}
\end{figure}

This section employs two exemplar scenarios, specifically multi-module hardware and state machine generation, to demonstrate the superior efficacy of multi-modal generation techniques over those that depend exclusively on natural language in the domain of structural hardware.
%画一个分层的图
\paragraph{Multi-Module Hardware Generation} Multi-module hardware architecting is prevalent in sophisticated hardware design paradigms. This case study investigates the potential superiority of multi-modal large-scale models over text-only language models in generating multi-module configurations. For instance, as illustrated in Fig. \ref{fig:nocpestructure}, a chain of Processing Elements (PEs) embodies a spatial architecture. We leverage OpenAI GPT4 Vision as the multi-modal large model to synthesize Verilog code. The findings depicted in Fig. \ref{fig:multimodulecase} demonstrate that within a multi-module context, the multi-modal large model proficiently captures intricate hardware details, thereby surpassing methods reliant solely on textual input. Consequently, text-only models are prone to inaccuracies and erroneous outputs.
\paragraph{State Machine Generation} State machines are frequently depicted using diagrams. As illustrated in Fig. \ref{fig:statemachineexample}, we utilize an image to represent the state machine responsible for detecting the sequence \texttt{10011}. When the input matches \texttt{10011}, the circuit outputs a logic high (1). OpenAI GPT4 Vision was selected as the multi-modal large model for this task. The generated Verilog code was assessed using the pass@5 criterion. The results demonstrated that the multi-modal model (GPT4V) successfully produced a version of the code that passed the testbench evaluations, whereas the text-only model version exhibited deficiencies in certain test cases. Fig. \ref{fig:casestudyresult} highlights the incorrect state transitions in red text. This example accentuates the superior capability of visual information in extracting structural details from images, thereby markedly enhancing the accuracy of code generation.


\subsection{Motivation: Do we need a new benchmark?}
Previous benchmarks such as RTLLLM and Verigen still have the following challenges in multi-modal generation environments.

\paragraph{Challenge I: Unstandardized Modal Input} Converting chip specifications to RTL code requires comprehension of the hardware diagram, which in turn depends on having a standardized form of the diagram representation. Rather than manually drafting a chip diagram, we suggest using a standardized diagram definition and query language to meet the specification's needs.

\paragraph{Challenge II: Lack of Complexity Categorization for Multi-Modal Input Prompts} Human developers frequently supply multi-modal model prompts with varying complexity across different levels of description, ranging from simple to elaborate. Nevertheless, existing benchmarks predominantly focus on prompts within natural-language-only contexts, thus failing to deliver exhaustive benchmarking outcomes.

\paragraph{Challenge III: Coarse-grain output metrics} Program completion plays an important role in EDA editor scenario. Most LLM uses the predict-the-next-token paradigm, which computes the loss of the next token. However, current benchmarks(\emph{e.g.} RTLLLM\cite{lu2023rtllm}, Verigen\cite{verigen}) directly provide holistic program output to compute their pass rate such as RTLLLM. Our benchmark splits the coarse-grain outputs into several fine-grain outputs, which provide a loss metric with the next $N$ tokens.

\section{Visual and Natural Language Hardware Co-Design Workflow}

%图中应该有query language的存在
\begin{figure}
    \centering
    \includegraphics[width=\linewidth]{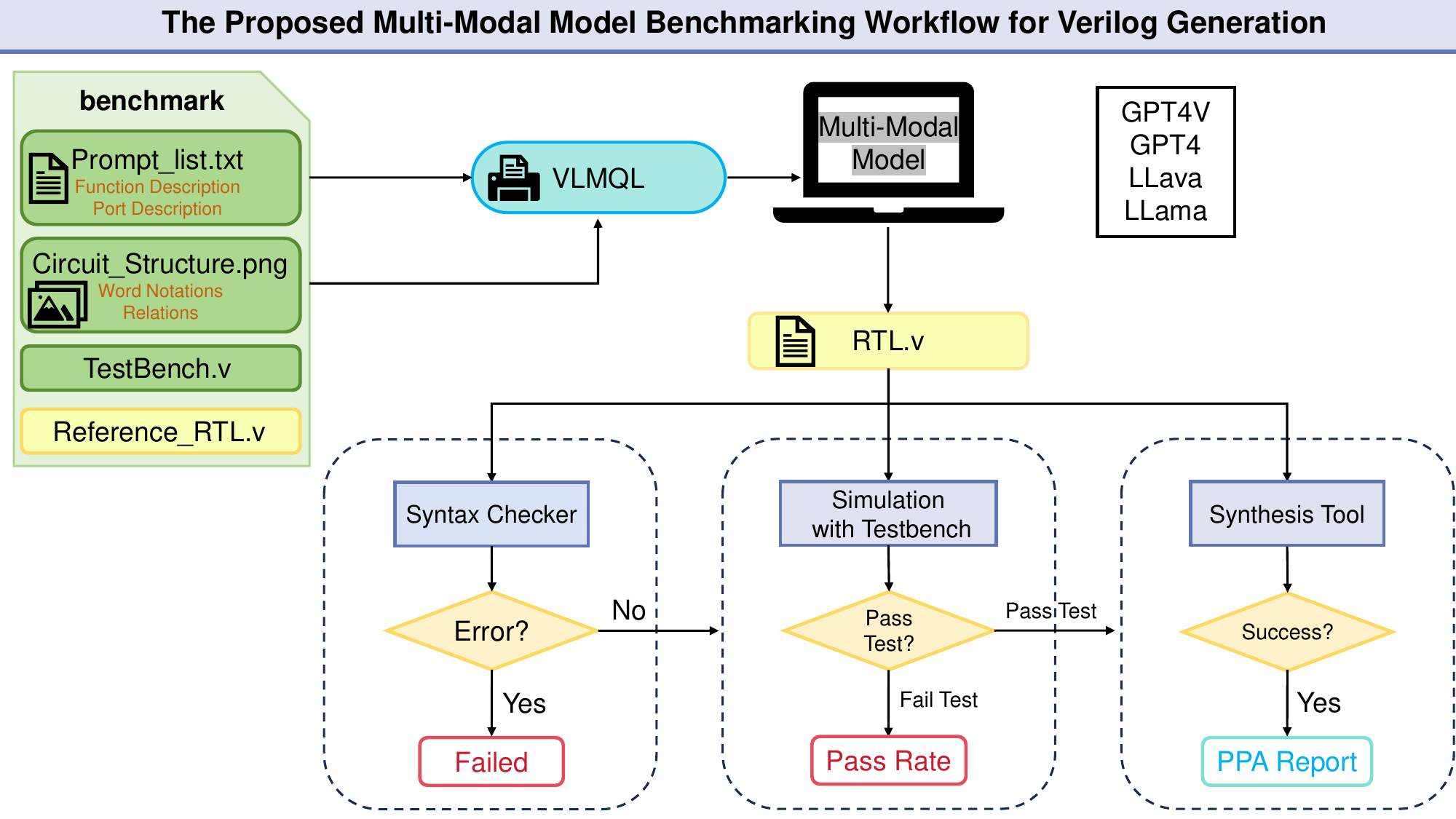}
    \caption{The overall end-to-end visual and natural language hardware co-generation workflow. The grey block is the proposed benchmark.}
    \label{fig:workflow}
\end{figure}
\subsection{Visual and Natural Language Co-Generation Workflow}  For an end-to-end Verilog generation flow, we split them as two parts as shown in Fig. \ref{fig:workflow}. The frontend accepts natural language and image(\emph{i.e.,} hardware structure) for Verilog generation and outputs Verilog  file.  The backend accepts the Verilog file and outputs the PPA reports, GDSII layout and function analysis report. To do a democratic hardware design, we chose siliconcompiler and openlane as the backend, which are open-source EDA tools for ASIC synthesis.

\subsection{The Formulation of the Circuit Structure in the Benchmark}\label{sec:codesign}

 To overcome the limitations of natural-language-only hardware generation and challenge I in Sec. \ref{sec:moti}, we illustrate the visual form and natural language form co-design knowledge as below. The visual form is used to generate top level connections and relations between each module. However, the detailed function can not be presented properly in the image/diagram only with visual information. The detailed information is collected in a natural language format. Several core concepts must be remembered in natural language designs below.  

\textbf{Word Notations} in the visual hardware graph: As shown in Fig. \ref{fig:nocpestructure},  Word Notations are the connections between visual and natural language hardware representation, which are the words on the image. Users declare the name on each block to illustrate it clearly in a natural language format. For example, for a five stage pipeline, the execution stage is annotated on the image and the natural language part can use "In the execution stage, the processor [DO SOMETHING]". Without word notations, large models only see the block, which can not detect the right description. Therefore, word notations are the interface between image design and natural language design.
%此处需要画图说明
\begin{itemize}
    \item Module name word notations: The diagram name provides a high-level name and description of the overall module or component represented visually. This allows connecting the diagram to natural language that references this module by name.
    \item Wire width word notations: Lines with variable width can represent connections of differing bandwidth or bitwidth. Annotating the width numerically clarifies the intent.
    \item Wire Function word notations: Text labels on wires indicate whether they carry data signals, control signals, clocks, etc. This clarifies the role of connections in the natural language.
    \item Block Function word notations: Major functional blocks are annotated with descriptive labels indicating their roles (\emph{e.g.,} ALU, multiplexer, register file). This allows precise natural language references.
    \item Ports word notations: Port labels designate the names and types of external interfaces. Natural language can then refer to interacting with specific ports.

    %Relations are represented by arrows and connections in the image, which constitute the core elements of a hardware structure
    
\end{itemize}
\paragraph{Definition: \textbf{Relations} in the visual hardware graph.} As shown in Fig. \ref{fig:nocpestructure}, relations are the arrows and connections on the image, which are the core parts of a hardware structure image. The relations in hardware structure connect the source and sink nodes, which represent inputs and outputs. According to relation theory in computer science, there are three possible relations on the image, one-to-one relations, one-to-many relations and many-to-many relations.
\begin{itemize}
    \item Single arrow relations: These represent singular connections between two components, like an output from one gate going to the input of another gate. They show a direct 1-to-1 relationship(\emph{e.g.,} wire).
    \item 1-to Many and Many-to-many arrow relations: These represent bus connections where multiple wires are bundled together. They model relationships where a group of outputs connects to a group of inputs, showing 1-to-many (\emph{e.g.,} bus) or many-to-many (\emph{e.g.,} crossbar) connectivity.
\end{itemize}

\paragraph{Definition: \textbf{Module Function Description} in natural language.} Module function description is the function in natural language format. For example, the sentence "a 3-8 decoder accepts a 3-bit number and outputs a 8-bit number where only one bit is one." is a typical module function description in natural language. The "3-8 decoder" is the word notation in the corresponding image.
\paragraph{Definition: \textbf{Module Port Description} in natural language.} Module port description is the port width and function description in natural language format. For example, the sentence "the input of 3-8 decoder is a 3-bit width wire named \texttt{innum}." is a typical port description in the natural language description.
%没有说明功能仅说明宽度，加入这是一个输入端口。

%fig2？
\subsection{The proposed Multi-modal Hardware Design Benchmark}%1页

 To select a benchmark scientifically, we form the benchmark selection as an optimization problem as Equ. \ref{equ:kldiverge}, where $B$ denotes the benchmark and $Data$ denotes world-wide data. The target of the benchmark is to be consistent with the worldwide data, which means that the difference between the distributions of the two datasets should be minimized. 
\begin{equation}\label{equ:kldiverge}
    \min D(B||Data)
\end{equation}

To implement Sec. \ref{sec:codesign}, we mitigate the presence of hardware image outliers lacking annotations through template coding, which transposes raw data into a standardized format as delineated in Sec. \ref{sec:codesign}. Specifically, template coding processes an annotation-free hardware module image, upon which annotations are appended in accordance with textual specifications. Moreover, we employ four-element pairs as the foundational units of the benchmark, comprising a textual modality (\emph{i.e.} \texttt{Prompt\_list.txt}), a diagrammatic modality (\emph{i.e.} \texttt{Circuit\_Structure.png}), a testbench for pass rate assessment (\emph{i.e.} \texttt{TestBench.v}), and a reference correct RTL Verilog program (\emph{i.e.} \texttt{reference.v}).
\paragraph{Benchmarking Output Complexity via Hierarchical Difficulty Workload} We categorize the workloads into three distinct levels, ranging from low to high complexity, to more effectively benchmark the performance of LLMs across varied design complexities, as illustrated in Table \ref{tab:benchmark}. Specifically, the arithmetic level encompasses fundamental numerical operations such as addition, multiplication, and division. The logic level includes standard controllers in hardware design, such as \texttt{edge\_detect} and \texttt{pulse\_detect}. The advanced level pertains to intricate units in CPU design (e.g., a 3-stage pipeline) and core units in matrix multiplication (e.g., 4$\times $4 GEMM).

\paragraph{Benchmarking Input Complexity via Multi-level Prompting} To address the challenge delineated in Sec. \ref{sec:moti}, we integrate multi-level prompts into the proposed multi-modal benchmark for pre-trained multi-modal models, thereby facilitating the assessment of design comprehension capabilities. For instance, an adept large language model (LLM) can accurately derive the RTL program from both rudimentary and intricate prompts. Specifically, the complexity of the prompts ranges from low to high. The low-level prompt primarily consists of a diagram devoid of detailed natural language exposition. The middle-level prompt includes a diagram accompanied by succinct natural language that outlines the core functionality. The high-level prompt furnishes exhaustive hardware information, encompassing details such as register specifications and clock edge information.

\paragraph{Fine-grain Output Measurement}  Driven by challenge III, we introduce a fine-grain output segmentation method aimed at providing more thorough evaluations through the use of token-by-token metrics. Assuming the inference result output is $\{tok'_0,tok'_1,\dots,tok'_N\}$ and the reference benchmark is $\{tok_0,tok_1,\dots,tok_N\}$, the fine-grain success metric is defined in Equation \ref{equ:lossequation}, where success is achieved when the tokens match.

\begin{equation}
\label{equ:lossequation}
    success^N =\Sigma_{i=0}^{N}1_{tok'_i = tok_i}
\end{equation}
\subsection{Facilitate Flexible Benchmarking using Query Language}
%注意用transformer的图表示他们之间的联系

\begin{figure}
    \centering
    \includegraphics[width=\linewidth]{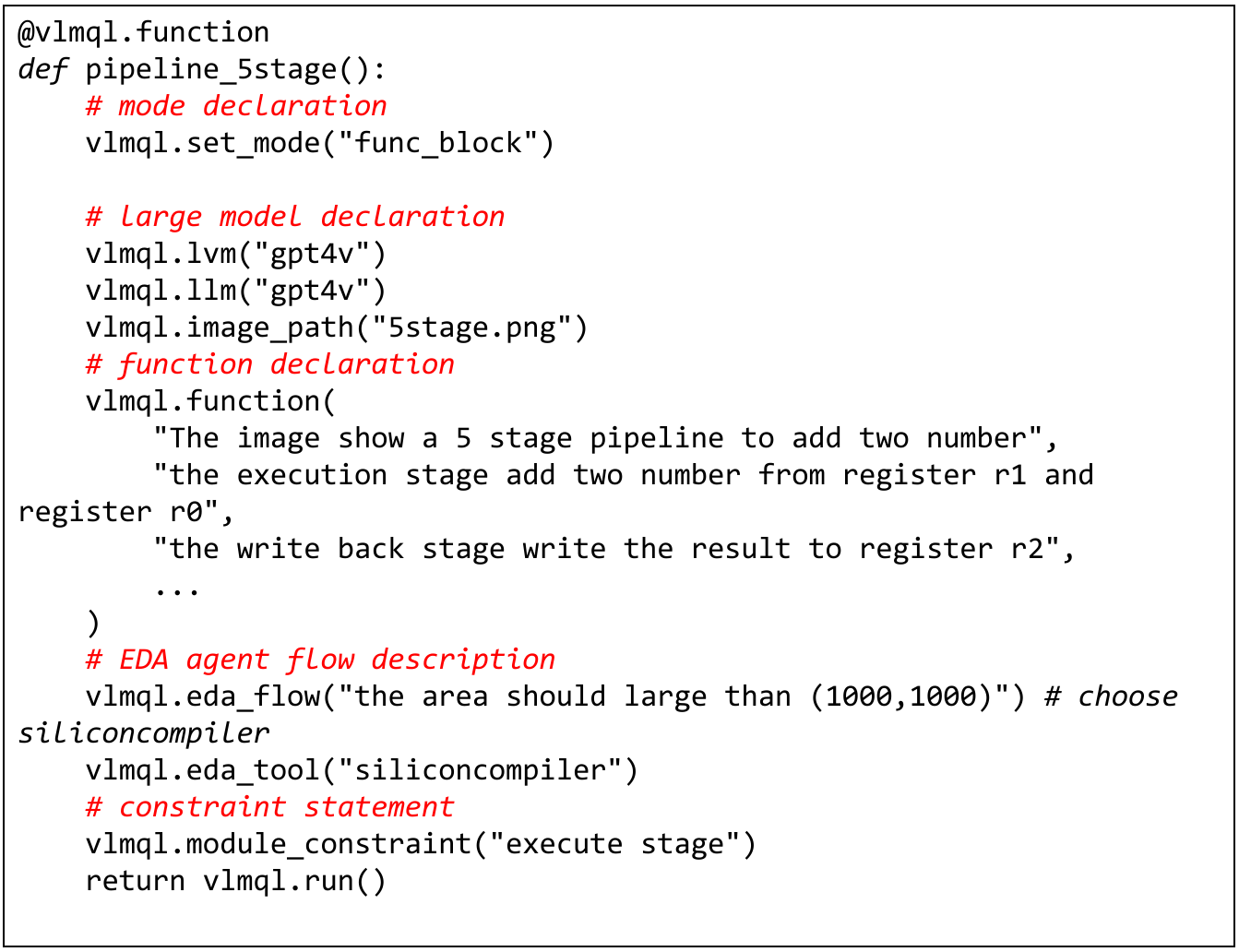}
    \caption{A case for VLMQL. The image shows a 5 stage pipeline, where we need to synthesis one of the stage in it.}
    \label{fig:vlmql}
\end{figure}

To execute the aforementioned visual and natural language co-design methodology, we introduce and realize a large model query language framework. By applying query language inference within a large language model, it is feasible to manage the LLM's output and input, thereby minimizing inference time and boosting accuracy through efficient network requests and template prompts\cite{lmql}.
We introduce a query language, named VLMQL (Verilog Large Model Query Language), specifically for generating Verilog using visual and natural language. 
\begin{table*}[]
\caption{Benchmark for multi-modal model Verilog generation.}
\label{tab:benchmark}
\resizebox{\linewidth}{!}{%
\begin{tabular}{|l|l|l|l|}
\hline
\multirow{2}{*}{\textbf{Type}} &
  \multirow{2}{*}{Name} &
  \multirow{2}{*}{Description} &
  \multirow{2}{*}{\begin{tabular}[c]{@{}l@{}}Approx. Code\\ Line\end{tabular}} \\
                             &                   &                                                                                                &     \\ \hline
\multirow{12}{*}{Advance} &
  1x4 systolic &
  A 1x4 systolic array with 1 row and 4 columns of processing elements for high throughput parallel processing. &
  18 \\ \cline{2-4} 
                             & 1x2 systolic      & A smaller 1x2 systolic array with 1 row and 2 columns of processing elements.                & 22  \\ \cline{2-4} 
                             & 2x2 systolic      & A 2x2 systolic array with 2 rows and 2 columns of processing   elements.                       & 11  \\ \cline{2-4} 
 &
  4x4 gemm &
  An array of 4x4 processing elements configured in a two-dimensional systolic structure for efficient and parallel matrix multiplication &  23 \\ \cline{2-4} 
 &
  5 stage pipeline & A 5-phase instruction pipeline that separates instruction processing into the stages: fetch, decode, execute, memory access, and writeback. &
  66 \\ \cline{2-4} 
     & 3 stage pipeline  & A simpler 3 stage instruction pipeline with fetch, execute and writeback stages.             & 39  \\ \cline{2-4} 
     & MAC PE            & A MAC (multiply-accumulate) processing element for performing vector matrix multiplications. & 12  \\ \cline{2-4} 
     & 2state\_fsm      & A state machine module to transition between 2 states based on inputs.                 & 23  \\ \cline{2-4} 
     & 3state\_fsm      & A state machine module to transition between 3 states   based on inputs.                 & 36  \\ \cline{2-4} 
     & 4state\_fsm      & A state machine module to transition between 4 states   based on inputs.                 & 38  \\ \cline{2-4} 
     & 5state\_fsm      & A state machine module to transition between 5 states   based on inputs.                 & 38  \\ \cline{2-4} 
     & 6state\_fsm      & A state machine module that detect the string 10011                & 77  \\ \hline

\multirow{8}{*}{Logic}       & Johnson\_Counter  & A 64-bit Johnson counter (torsional ring counter)                                              & 12  \\ \cline{2-4} 
                             & alu               & An ALU for 32bit MIPS-ISA CPU                                                                  & 99  \\ \cline{2-4} 
                             & edge\_detect      & A module for edge   detection, there is a slowly changing 1 bit signal a.                      & 35  \\ \cline{2-4} 
                             & freq\_div         & A frequency divider that the input clock frequency of 100MHz   signal,                         & 46  \\ \cline{2-4} 
                             & mux               & A multi-bit MUX synchronizer                                                                   & 40  \\ \cline{2-4} 
                             & parallel2serial   & A module for parallel-to-serial conversion                                                     & 33  \\ \cline{2-4} 
                             & pulse\_detect     & A module for edge detection, there is a slowly changing 1 bit   signal data\_in.               & 29  \\ \cline{2-4} 
                             & right\_shifter    & A right shifter                                                                                & 12  \\ \hline
\multirow{10}{*}{Arithmetic} & accu              & A module to achieve serial input data accumulation output.                                     & 51  \\ \cline{2-4} 
                             & adder\_16bit      & A module of a 16-bit full adder.                                                               & 102 \\ \cline{2-4} 
                             & add\_16bit\_csa   & A 16-bit carry select adder.                                                                   & 114 \\ \cline{2-4} 
                             & adder\_32bit      & A module of a carry lookahead 32 bit adder based on CLAs.                                      & 152 \\ \cline{2-4} 
                             & adder\_64\_bit    & A module of a ripple 64 bit adder.                                                             & 171 \\ \cline{2-4} 
                             & adder\_8bit       & A module of an 8 bit adder in gate level.                                                      & 20  \\ \cline{2-4} 
                             & div\_16bit        & A 16-bit divider module, dividend is 16-bit and divider is   16-bit.                           & 33  \\ \cline{2-4} 
                             & multi\_booth      & An 8bit booth-4 multiplier                                                                     & 71  \\ \cline{2-4} 
                             & multi\_pipe\_4bit & The design of 4bit unsigned number pipeline multiplier.                                        & 33  \\ \cline{2-4} 
                             & multipipe\_8bit   & The design of unsigned 8bit multiplier based on pipelining   processing.                       & 76  \\ \hline
\end{tabular}%
}
\end{table*}
\paragraph{VLMQL Framework} The Visual and Language Model Query Language (VLMQL) is meticulously crafted to encapsulate the dual modalities of visual and textual hardware co-design, as delineated in Section \ref{sec:codesign}. Functioning as an advanced form of controllable prompt engineering, the cornerstone of VLMQL resides in a Python function illustrated in Figure \ref{fig:vlmql}. This function's output serves as the input for high-capacity vision-language models. The VLMQL framework is systematically partitioned into three integral components: declarations for visual and natural language inputs, a detailed agent flow schema for Electronic Design Automation (EDA) tool operations, and constraint formulations for prompt scheduling.
\paragraph{Mode Declaration: Three-tier hardware representation} The hardware visual description is divided into three tiers: gate-level, algorithm-level, and function block-level, ranging from concrete to abstract. Gate-level depicts the image as logic gates. Algorithm-level encompasses the basic blocks in the image, which are the familiar algorithms in large models, such as add and multiply blocks. For larger and customized designs, VLMQL employs the function block level to represent these elements, indicating that these blocks require further illustration. These levels are declared as the parameter as \texttt{vlmql.set\_mode("func\_block")}, which is the prompt of \textbf{Module function description}. The declaration establishes a flexible guideline for users to create the diagram, which is then converted into a natural language prompt, like "The basic block of the input image is a logic gate.".
\paragraph{Large Model Declaration: Model Parameters and Their Inputs.} For large models, declarations specify the task to be taken as input and then generate a prompt or select the function inputs.
\begin{itemize}
    \item Model Selection: To choose the target vision model and large language model, VLMQL uses \texttt{vlmql.lvm("[vision model]")} and \texttt{vlmql.llm("[language model]")} to choose the target model.
    \item Image Selection: To choose the input hardware structure image, VLMQL uses \texttt{vlmql.image\_path("[image path]")}, which can obtain images and feed them into the large language model. 

\end{itemize}

% \zrc{\paragraph{Image Drawing Tool: To facilitate drawing the design image, we design a chip image drawing tool in VLMQL. The workflow of the drawing tool is 1. Set the name of the whole design (\emph{i.e.} adder\_8bit); 2. Define the sub-modules in the design; For each submodule, we need to define the name of the sub-module and the ports of the sub-module (\emph{i.e.} full\_adder; input a, b, cin; output sum, cout); 3. Define the connections among submodules (\emph{i.e.} full\_adder1:cout, full\_adder2:cin)}; 4. Use the graphviz tool to draw the chip diagram.}

\paragraph{Function Declaration: Module Separation Description.} The function declarations consist of a sequence of prompts. Each line in the function acts as a prompt adhering to the specification in Sec. \ref{sec:codesign}. The initial section provides the function description for each Verilog module, while the latter part covers the port description for every Verilog module.
\paragraph{EDA Agent Flow Description.} In addition to using large models to generate Verilog files, generating EDA scripts is equally crucial\cite{he2023chateda}. VLMQL supports the description of a complete visual and natural language co-design workflow.
\begin{itemize}
    \item Tool selection: These parts select the EDA tool as \texttt{vlmql. eda\_tool ( "[EDATool]" )}. The Large model might not be familiar with EDA tool scripting. Thus, the VLMQL compiler offers an EDA tool script example as a prompt for the LLM, a process known as in-context learning.
    \item EDA Flow detailing: This section communicates intricate details to the LLM using natural language. For instance, siliconcompiler needs the input file name as well as the floorplan dimensions. This acts as a prompt for the LLM, using the primitive \texttt{vlmql.eda\_flow("[natural language description]")}. An example of a "natural language description" could be "modify the overall area to 210*210".
\end{itemize}

\paragraph{Constraint Statement: Constrain the input and output for low-cost generation.} For some scenes, users do not need to obtain all of the Verilog code of the module. For example, users may input a 5-stage CPU description prompt, and users only want to get the Verilog code of the execution stage. However, if following the typical workflow, the additional module except execution may cost many tokens.  Therefore, constraints are necessary to release these additional tokens. These constraints are finally compiled into prompts. For example,  For Verilog generation task, users can use \texttt{vlmql.module\_constraint("[modulename]")} to declare the module they want to generate, while other modules won't be generated. These constraints will serve as a guide.

\section{Evaluation}
% \usepackage{multirow}
% \usepackage{graphicx}
\begin{table*}[]
\caption{Syntax represents the number of Verilog code generated by LLM with syntax errors under pass@5. Function represents
the testbench pass rate of the best-performing Verilog code under pass@5. V represents Vision modal and T represents natural language text modal.}
\label{tab:evalgeneration}
\resizebox{\textwidth}{!}{%
\begin{tabular}{|cl|cc|cc|cc|cc|cc|cc|}
\hline
\multicolumn{2}{|c|}{\multirow{2}{*}{benchmark}} & \multicolumn{2}{c|}{GPT4-V   (V+T)} & \multicolumn{2}{c|}{GPT4   (T)} & \multicolumn{2}{c|}{GPT4-V (V)} & \multicolumn{2}{c|}{LLaVa (V+T)} & \multicolumn{2}{c|}{Llama   (T)} & \multicolumn{2}{c|}{LLaVa(V)} \\ \cline{3-14} 
\multicolumn{2}{|c|}{} & \multicolumn{1}{c|}{syntax} & function & \multicolumn{1}{c|}{syntax} & function & \multicolumn{1}{c|}{syntax} & function & \multicolumn{1}{c|}{syntax} & function & \multicolumn{1}{c|}{syntax} & function & \multicolumn{1}{c|}{syntax} & function \\ \hline
\multicolumn{1}{|c|}{\multirow{10}{*}{Logic}} & Johnson\_Counter & \multicolumn{1}{c|}{0} & 98\% & \multicolumn{1}{c|}{0} & 100\% & \multicolumn{1}{c|}{0} & 98\% & \multicolumn{1}{c|}{5} & 0\% & \multicolumn{1}{c|}{0} & 97\% & \multicolumn{1}{c|}{1} & 0\% \\ \cline{2-14} 
\multicolumn{1}{|c|}{} & alu & \multicolumn{1}{c|}{0} & 100\% & \multicolumn{1}{c|}{0} & 0\% & \multicolumn{1}{c|}{1} & 0\% & \multicolumn{1}{c|}{5} & 0\% & \multicolumn{1}{c|}{5} & 0\% & \multicolumn{1}{c|}{1} & 0\% \\ \cline{2-14} 
\multicolumn{1}{|c|}{} & edge\_detect & \multicolumn{1}{c|}{0} & 100\% & \multicolumn{1}{c|}{0} & 100\% & \multicolumn{1}{c|}{0} & 100\% & \multicolumn{1}{c|}{0} & 98\% & \multicolumn{1}{c|}{0} & 98\% & \multicolumn{1}{c|}{0} & 0\% \\ \cline{2-14} 
\multicolumn{1}{|c|}{} & freq\_div & \multicolumn{1}{c|}{0} & 100\% & \multicolumn{1}{c|}{0} & 0\% & \multicolumn{1}{c|}{1} & 0\% & \multicolumn{1}{c|}{4} & 0\% & \multicolumn{1}{c|}{5} & 0\% & \multicolumn{1}{c|}{1} & 0\% \\ \cline{2-14} 
\multicolumn{1}{|c|}{} & mux & \multicolumn{1}{c|}{0} & 100\% & \multicolumn{1}{c|}{1} & 100\% & \multicolumn{1}{c|}{0} & 10\% & \multicolumn{1}{c|}{1} & 0\% & \multicolumn{1}{c|}{5} & 0\% & \multicolumn{1}{c|}{0} & 100\% \\ \cline{2-14} 
\multicolumn{1}{|c|}{} & parallel2serial & \multicolumn{1}{c|}{0} & 0\% & \multicolumn{1}{c|}{0} & 0\% & \multicolumn{1}{c|}{1} & 0\% & \multicolumn{1}{c|}{0} & 100\% & \multicolumn{1}{c|}{5} & 0\% & \multicolumn{1}{c|}{0} & 0\% \\ \cline{2-14} 
\multicolumn{1}{|c|}{} & pulse\_detect & \multicolumn{1}{c|}{0} & 100\% & \multicolumn{1}{c|}{0} & 0\% & \multicolumn{1}{c|}{5} & 0\% & \multicolumn{1}{c|}{0} & 0\% & \multicolumn{1}{c|}{5} & 0\% & \multicolumn{1}{c|}{3} & 0\% \\ \cline{2-14} 
\multicolumn{1}{|c|}{} & right\_shifter & \multicolumn{1}{c|}{0} & 100\% & \multicolumn{1}{c|}{0} & 100\% & \multicolumn{1}{c|}{0} & 100\% & \multicolumn{1}{c|}{0} & 100\% & \multicolumn{1}{c|}{0} & 100\% & \multicolumn{1}{c|}{0} & 0\% \\ \cline{2-14} 
\multicolumn{1}{|c|}{} & serial2parallel & \multicolumn{1}{c|}{0} & 100\% & \multicolumn{1}{c|}{0} & 100\% & \multicolumn{1}{c|}{2} & 0\% & \multicolumn{1}{c|}{5} & 0\% & \multicolumn{1}{c|}{0} & 0\% & \multicolumn{1}{c|}{1} & 0\% \\ \cline{2-14} 
\multicolumn{1}{|c|}{} & width\_8to16 & \multicolumn{1}{c|}{0} & 100\% & \multicolumn{1}{c|}{0} & 100\% & \multicolumn{1}{c|}{0} & 100\% & \multicolumn{1}{c|}{1} & 0\% & \multicolumn{1}{c|}{0} & 34\% & \multicolumn{1}{c|}{1} & 0\% \\ \hline
\multicolumn{1}{|c|}{\multirow{10}{*}{Arithmetic}} & accu & \multicolumn{1}{c|}{0} & 100\% & \multicolumn{1}{c|}{0} & 100\% & \multicolumn{1}{c|}{3} & 0\% & \multicolumn{1}{c|}{5} & 0\% & \multicolumn{1}{c|}{5} & 0\% & \multicolumn{1}{c|}{3} & 0\% \\ \cline{2-14} 
\multicolumn{1}{|c|}{} & adder\_16bit & \multicolumn{1}{c|}{0} & 100\% & \multicolumn{1}{c|}{0} & 100\% & \multicolumn{1}{c|}{0} & 100\% & \multicolumn{1}{c|}{5} & 0\% & \multicolumn{1}{c|}{5} & 0\% & \multicolumn{1}{c|}{8} & 0\% \\ \cline{2-14} 
\multicolumn{1}{|c|}{} & add\_16bit\_csa & \multicolumn{1}{c|}{0} & 100\% & \multicolumn{1}{c|}{0} & 100\% & \multicolumn{1}{c|}{5} & 0\% & \multicolumn{1}{c|}{2} & 100\% & \multicolumn{1}{c|}{1} & 100\% & \multicolumn{1}{c|}{2} & 0\% \\ \cline{2-14} 
\multicolumn{1}{|c|}{} & adder\_32bit & \multicolumn{1}{c|}{3} & 0\% & \multicolumn{1}{c|}{0} & 0\% & \multicolumn{1}{c|}{1} & 0\% & \multicolumn{1}{c|}{5} & 0\% & \multicolumn{1}{c|}{5} & 0\% & \multicolumn{1}{c|}{1} & 0\% \\ \cline{2-14} 
\multicolumn{1}{|c|}{} & adder\_64\_bit & \multicolumn{1}{c|}{0} & 0\% & \multicolumn{1}{c|}{2} & 0\% & \multicolumn{1}{c|}{1} & 0\% & \multicolumn{1}{c|}{5} & 0\% & \multicolumn{1}{c|}{5} & 0\% & \multicolumn{1}{c|}{4} & 0\% \\ \cline{2-14} 
\multicolumn{1}{|c|}{} & adder\_8bit & \multicolumn{1}{c|}{0} & 100\% & \multicolumn{1}{c|}{0} & 100\% & \multicolumn{1}{c|}{2} & 0\% & \multicolumn{1}{c|}{0} & 0\% & \multicolumn{1}{c|}{3} & 0\% & \multicolumn{1}{c|}{0} & 0\% \\ \cline{2-14} 
\multicolumn{1}{|c|}{} & div\_16bit & \multicolumn{1}{c|}{1} & 0\% & \multicolumn{1}{c|}{5} & 0\% & \multicolumn{1}{c|}{1} & 0\% & \multicolumn{1}{c|}{5} & 0\% & \multicolumn{1}{c|}{5} & 0\% & \multicolumn{1}{c|}{2} & 0\% \\ \cline{2-14} 
\multicolumn{1}{|c|}{} & multi\_booth & \multicolumn{1}{c|}{0} & 50\% & \multicolumn{1}{c|}{1} & 0\% & \multicolumn{1}{c|}{1} & 0\% & \multicolumn{1}{c|}{2} & 0\% & \multicolumn{1}{c|}{5} & 0\% & \multicolumn{1}{c|}{1} & 0\% \\ \cline{2-14} 
\multicolumn{1}{|c|}{} & multi\_pipe\_4bit & \multicolumn{1}{c|}{0} & 50\% & \multicolumn{1}{c|}{3} & 100\% & \multicolumn{1}{c|}{2} & 0\% & \multicolumn{1}{c|}{4} & 100\% & \multicolumn{1}{c|}{1} & 0\% & \multicolumn{1}{c|}{5} & 0\% \\ \cline{2-14} 
\multicolumn{1}{|c|}{} & multipipe\_8bit & \multicolumn{1}{c|}{0} & 100\% & \multicolumn{1}{c|}{1} & 0\% & \multicolumn{1}{c|}{1} & 0\% & \multicolumn{1}{c|}{5} & 0\% & \multicolumn{1}{c|}{5} & 0\% & \multicolumn{1}{c|}{0} & 0\% \\ \hline
\multicolumn{1}{|c|}{\multirow{12}{*}{Advanced}} & 1x2nocpe & \multicolumn{1}{c|}{0} & 100\% & \multicolumn{1}{c|}{3} & 0\% & \multicolumn{1}{c|}{3} & 0\% & \multicolumn{1}{c|}{0} & 34\% & \multicolumn{1}{c|}{4} & 0\% & \multicolumn{1}{c|}{0} & 0\% \\ \cline{2-14} 
\multicolumn{1}{|c|}{} & 1x4systolic & \multicolumn{1}{c|}{0} & 100\% & \multicolumn{1}{c|}{0} & 100\% & \multicolumn{1}{c|}{1} & 0\% & \multicolumn{1}{c|}{5} & 0\% & \multicolumn{1}{c|}{5} & 0\% & \multicolumn{1}{c|}{1} & 0\% \\ \cline{2-14} 
\multicolumn{1}{|c|}{} & 2x2systolic & \multicolumn{1}{c|}{0} & 100\% & \multicolumn{1}{c|}{2} & 0\% & \multicolumn{1}{c|}{5} & 0\% & \multicolumn{1}{c|}{5} & 0\% & \multicolumn{1}{c|}{5} & 0\% & \multicolumn{1}{c|}{2} & 0\% \\ \cline{2-14} 
\multicolumn{1}{|c|}{} & 3stagepipe & \multicolumn{1}{c|}{1} & 0\% & \multicolumn{1}{c|}{5} & 0\% & \multicolumn{1}{c|}{5} & 0\% & \multicolumn{1}{c|}{5} & 0\% & \multicolumn{1}{c|}{5} & 0\% & \multicolumn{1}{c|}{1} & 0\% \\ \cline{2-14} 
\multicolumn{1}{|c|}{} & 4x4spatialacc & \multicolumn{1}{c|}{3} & 0\% & \multicolumn{1}{c|}{5} & 0\% & \multicolumn{1}{c|}{5} & 0\% & \multicolumn{1}{c|}{5} & 0\% & \multicolumn{1}{c|}{5} & 0\% & \multicolumn{1}{c|}{1} & 0\% \\ \cline{2-14} 
\multicolumn{1}{|c|}{} & 5stagepipe & \multicolumn{1}{c|}{5} & 0\% & \multicolumn{1}{c|}{5} & 0\% & \multicolumn{1}{c|}{5} & 0\% & \multicolumn{1}{c|}{5} & 0\% & \multicolumn{1}{c|}{5} & 0\% & \multicolumn{1}{c|}{1} & 0\% \\ \cline{2-14} 
\multicolumn{1}{|c|}{} & fsm & \multicolumn{1}{c|}{0} & 100\% & \multicolumn{1}{c|}{0} & 0\% & \multicolumn{1}{c|}{1} & 0\% & \multicolumn{1}{c|}{5} & 0\% & \multicolumn{1}{c|}{3} & 0\% & \multicolumn{1}{c|}{10} & 0\% \\ \cline{2-14} 
\multicolumn{1}{|c|}{} & macpe & \multicolumn{1}{c|}{0} & 100\% & \multicolumn{1}{c|}{0} & 100\% & \multicolumn{1}{c|}{3} & 0\% & \multicolumn{1}{c|}{0} & 100\% & \multicolumn{1}{c|}{2} & 0\% & \multicolumn{1}{c|}{2} & 0\% \\ \cline{2-14} 
\multicolumn{1}{|c|}{} & statemachine & \multicolumn{1}{c|}{0} & 100\% & \multicolumn{1}{c|}{0} & 0\% & \multicolumn{1}{c|}{5} & 0\% & \multicolumn{1}{c|}{5} & 0\% & \multicolumn{1}{c|}{3} & 0\% & \multicolumn{1}{c|}{1} & 0\% \\ \cline{2-14} 
\multicolumn{1}{|c|}{} & 5state\_fsm & \multicolumn{1}{c|}{0} & 100\% & \multicolumn{1}{c|}{0} & 0\% & \multicolumn{1}{c|}{3} & 0\% & \multicolumn{1}{c|}{5} & 0\% & \multicolumn{1}{c|}{5} & 0\% & \multicolumn{1}{c|}{3} & 0\% \\ \cline{2-14} 
\multicolumn{1}{|c|}{} & 3state\_fsm & \multicolumn{1}{c|}{0} & 100\% & \multicolumn{1}{c|}{0} & 100\% & \multicolumn{1}{c|}{0} & 100\% & \multicolumn{1}{c|}{5} & 0\% & \multicolumn{1}{c|}{4} & 0\% & \multicolumn{1}{c|}{0} & 0\% \\ \cline{2-14} 
\multicolumn{1}{|c|}{} & 4state\_fsm & \multicolumn{1}{c|}{0} & 0\% & \multicolumn{1}{c|}{0} & 100\% & \multicolumn{1}{c|}{0} & 100\% & \multicolumn{1}{c|}{3} & 100\% & \multicolumn{1}{c|}{5} & 0\% & \multicolumn{1}{c|}{1} & 0\% \\ \hline
\multicolumn{2}{|c|}{Success Rate} & \multicolumn{1}{c|}{84.38\%} & 71.81\% & \multicolumn{1}{c|}{68.75\%} & 46.88\% & \multicolumn{1}{c|}{25.00\%} & 33.90\% & \multicolumn{1}{c|}{34.38\%} & 25.88\% & \multicolumn{1}{c|}{21.88\%} & 13.41\% & \multicolumn{1}{c|}{25.00\%} & 3.13\% \\ \hline
\end{tabular}%
}
\end{table*}
\subsection{Evaluation Setup}
% \paragraph{Benchmark}
% \paragraph{Environment} VLMQL is implemented with LMQL\cite{lmql} in Python. 
% \paragraph{Baseline} Ours method uses the image with . We use the same natural language input without image as the baseline.

Our study systematically explores the efficacy of multi-modal language models in generating Verilog code by establishing a comprehensive benchmark that spans a spectrum of complexity across three categories: arithmetic, digital circuit, and advanced hardware designs. The benchmark is meticulously structured to assess the incremental benefits of multi-modal inputs as we progress from simple to more complex cases. To evaluate the performance, each model within the GPT-4 and LLaMA series is tasked with generating Verilog code for each category five times, facilitating a robust and iterative testing methodology.

The evaluation focuses on three critical aspects: syntax correctness, functional accuracy, and next-token accuracy, ensuring a holistic assessment of the Verilog code generated. The benchmark leverages a "pass@5" metric, which examines the best Verilog code within five attempts, providing insight into both the precision and reliability of the models.

\subsection{Evaluation Result}
% \usepackage{multirow}
% \usepackage{graphicx}

The evaluation of multi-modal language models in the generation of Verilog code has yielded illuminating results, particularly in terms of syntax and functionality correctness.

\paragraph{Syntax Correctness}
The introduction of vision representations has led to a notable enhancement in syntax correctness compared with natural-language-only representation. Specifically,  as shown in Tab. \ref{tab:evalgeneration}, when adding vision modal, the GPT-4 series show an increase in syntax success rate from 68.75\% to 84.38\%. The LLaMA series also demonstrated an improvement in syntax success rate from 21.88\% to 34.38\%.  We speculate that this improvement highlights the models' increase proficiency in understanding  syntactic rules of Verilog with the aid of visual context. In addition, these results suggest that while all models benefit from multi-modal inputs, those with more advanced architectures, like the GPT-4 series, exhibit a greater propensity for minimizing syntax errors.

\paragraph{Functionality Correctness}
Considering functional correctness, the passing rate has significant improvement when adding vision modal.  Specifically, as shown in Tab. \ref{tab:evalgeneration}, for the GPT-4 series, the success rate improves remarkably from 46.88\% to 71.81\% with the integration of visual inputs. The LLaMA series, conversely, experienced a modest improvement, with pass rates increasing from 13.41\% to 25.88\%. This underscores the models' enhanced capability to not only generate syntactically correct code but also functionally correct Verilog.  Besides, while the starting point for functionality correctness was lower for the LLaMA series, the addition of multi-modal data still contributed to a meaningful improvement in performance.

%The results collectively indicate that multi-modality significantly uplifts the language models' ability to generate Verilog code that is both syntactically and functionally more accurate. These advancements attest to the substantial potential of incorporating visual data to augment the efficacy of language models in complex tasks like hardware design, ensuring that the code generated not only looks correct but also works correctly in practical applications.

\subsection{Sensitivity Study} 
\paragraph{Multi-Level Multi-Modal Prompt} We benchmark the success rate on multi-level prompts from simple to complex as shown in Tab. \ref{tab:promptsensi}. The results show that with the increase of the prompt information, the success rate has a further increase in most cases. Specifically, the prompts 
 change from simple to detailed, while the success rate ranges from 40.63\% to 71.81\% in GPT4-V, from 9.38\% to 25.88\% in LLaVa. Therefore, the proposed multi-level prompt can distinguish the LLM-generating difference.

\paragraph{Fine-grain output Prompt} In addition, we measure the output verilog program with the output metric as shown in Tab. \ref{tab:nexttoken}, which compares the next token prediction success rate, reflecting the LLM's program completion ability. The results show that the natural-language-only mode is weaker than the natural language and image co-design mode. Specifically, compared to the natural-language-only mode, the average success rate of the co-design mode improves from 63.64\% to 71.72\% in GPT series, from 20.20\% to 28.28\% in LLaMa series. These results support our speculation that the co-design mode is better than the natural-language-only mode in the token prediction task.

\paragraph{State Number Changes in FSM} To further explore the LLM sensitivity to design complexity, we analyze several control modules with state number changing (\emph{i.e.} push button LED) in Tab. \ref{tab:statechange} with GPT4-V as the base model, where the transition in the table denotes the state transition success rate, the state denotes the state register declaration correctness, the output denotes the signal output correctness in every state.  The results show that with the state number increase from 2 to 9, the success rate decreases from 100\% to 0\%. This shows that current LLM-generated hardware can not well capture the long distance information.

\begin{table}[htbp]
\caption{The state number change success rate on push button case. This shows the generating success rate in pass@5.}
\label{tab:statechange}
\scalebox{0.85}{
\begin{tabular}{|l|l|l|l|l|l|l|l|l|}
\hline
\textbf{\begin{tabular}[c]{@{}l@{}}State\\ Number\end{tabular}} & 2     & 3     & 4     & 5     & 6     & 7     & 8     & 9     \\ \hline
Transition                                                      & 100\% & 100\% & 80\%  & 60\%  & 60\%  & 20\%  & 0\%   & 0\%   \\ \hline
State                                                           & 100\% & 100\% & 100\% & 100\% & 100\% & 100\% & 100\% & 100\% \\ \hline
Output                                                          & 100\% & 100\% & 100\% & 100\% & 100\% & 100\% & 80\%  & 80\%  \\ \hline
\end{tabular}
}
\end{table}

% \usepackage{multirow}
% \usepackage{graphicx}
\begin{table}[htbp]
\caption{The success rate changes from simple to complex. S,M,C represent simple, medium and complex prompts respectively.}
\label{tab:promptsensi}
\resizebox{\linewidth}{!}{%
\begin{tabular}{|cl|rrr|rrr|}
\hline
\multicolumn{2}{|c|}{\multirow{2}{*}{\textbf{\begin{tabular}[c]{@{}c@{}}Benchmark\\ (function)\end{tabular}}}} & \multicolumn{3}{c|}{\textbf{LLaVa}} & \multicolumn{3}{c|}{\textbf{GPT4-V}} \\ \cline{3-8} 
\multicolumn{2}{|c|}{} & \multicolumn{1}{c|}{\textbf{S}} & \multicolumn{1}{c|}{\textbf{M}} & \multicolumn{1}{c|}{\textbf{C}} & \multicolumn{1}{c|}{\textbf{S}} & \multicolumn{1}{c|}{\textbf{M}} & \multicolumn{1}{c|}{\textbf{C}} \\ \hline
\multicolumn{1}{|c|}{\multirow{10}{*}{Logic}} & Johnson\_Counter & \multicolumn{1}{r|}{0\%} & \multicolumn{1}{r|}{0\%} & 0\% & \multicolumn{1}{r|}{0\%} & \multicolumn{1}{r|}{0\%} & 98\% \\ \cline{2-8} 
\multicolumn{1}{|c|}{} & alu & \multicolumn{1}{r|}{0\%} & \multicolumn{1}{r|}{0\%} & 0\% & \multicolumn{1}{r|}{20\%} & \multicolumn{1}{r|}{0\%} & 100\% \\ \cline{2-8} 
\multicolumn{1}{|c|}{} & edge\_detect & \multicolumn{1}{r|}{0\%} & \multicolumn{1}{r|}{40\%} & 98\% & \multicolumn{1}{r|}{100\%} & \multicolumn{1}{r|}{100\%} & 100\% \\ \cline{2-8} 
\multicolumn{1}{|c|}{} & freq\_div & \multicolumn{1}{r|}{0\%} & \multicolumn{1}{r|}{0\%} & 96\% & \multicolumn{1}{r|}{100\%} & \multicolumn{1}{r|}{80\%} & 100\% \\ \cline{2-8} 
\multicolumn{1}{|c|}{} & mux & \multicolumn{1}{r|}{0\%} & \multicolumn{1}{r|}{0\%} & 0\% & \multicolumn{1}{r|}{80\%} & \multicolumn{1}{r|}{100\%} & 100\% \\ \cline{2-8} 
\multicolumn{1}{|c|}{} & parallel2serial & \multicolumn{1}{r|}{0\%} & \multicolumn{1}{r|}{0\%} & 100\% & \multicolumn{1}{r|}{0\%} & \multicolumn{1}{r|}{0\%} & 0\% \\ \cline{2-8} 
\multicolumn{1}{|c|}{} & pulse\_detect & \multicolumn{1}{r|}{0\%} & \multicolumn{1}{r|}{0\%} & 0\% & \multicolumn{1}{r|}{100\%} & \multicolumn{1}{r|}{100\%} & 100\% \\ \cline{2-8} 
\multicolumn{1}{|c|}{} & right\_shifter & \multicolumn{1}{r|}{0\%} & \multicolumn{1}{r|}{100\%} & 100\% & \multicolumn{1}{r|}{100\%} & \multicolumn{1}{r|}{100\%} & 100\% \\ \cline{2-8} 
\multicolumn{1}{|c|}{} & serial2parallel & \multicolumn{1}{r|}{0\%} & \multicolumn{1}{r|}{0\%} & 0\% & \multicolumn{1}{r|}{0\%} & \multicolumn{1}{r|}{100\%} & 100\% \\ \cline{2-8} 
\multicolumn{1}{|c|}{} & width\_8to16 & \multicolumn{1}{r|}{0\%} & \multicolumn{1}{r|}{0\%} & 0\% & \multicolumn{1}{r|}{100\%} & \multicolumn{1}{r|}{100\%} & 100\% \\ \hline
\multicolumn{1}{|c|}{\multirow{10}{*}{Arithmetic}} & accu & \multicolumn{1}{r|}{0\%} & \multicolumn{1}{r|}{0\%} & 0\% & \multicolumn{1}{r|}{0\%} & \multicolumn{1}{r|}{0\%} & 100\% \\ \cline{2-8} 
\multicolumn{1}{|c|}{} & adder\_16bit & \multicolumn{1}{r|}{0\%} & \multicolumn{1}{r|}{40\%} & 0\% & \multicolumn{1}{r|}{100\%} & \multicolumn{1}{r|}{100\%} & 100\% \\ \cline{2-8} 
\multicolumn{1}{|c|}{} & add\_16bit\_csa & \multicolumn{1}{r|}{0\%} & \multicolumn{1}{r|}{0\%} & 100\% & \multicolumn{1}{r|}{0\%} & \multicolumn{1}{r|}{100\%} & 100\% \\ \cline{2-8} 
\multicolumn{1}{|c|}{} & adder\_32bit & \multicolumn{1}{r|}{0\%} & \multicolumn{1}{r|}{0\%} & 0\% & \multicolumn{1}{r|}{0\%} & \multicolumn{1}{r|}{0\%} & 0\% \\ \cline{2-8} 
\multicolumn{1}{|c|}{} & adder\_64\_bit & \multicolumn{1}{r|}{0\%} & \multicolumn{1}{r|}{0\%} & 0\% & \multicolumn{1}{r|}{0\%} & \multicolumn{1}{r|}{0\%} & 0\% \\ \cline{2-8} 
\multicolumn{1}{|c|}{} & adder\_8bit & \multicolumn{1}{r|}{100\%} & \multicolumn{1}{r|}{100\%} & 0\% & \multicolumn{1}{r|}{100\%} & \multicolumn{1}{r|}{100\%} & 100\% \\ \cline{2-8} 
\multicolumn{1}{|c|}{} & div\_16bit & \multicolumn{1}{r|}{0\%} & \multicolumn{1}{r|}{0\%} & 0\% & \multicolumn{1}{r|}{0\%} & \multicolumn{1}{r|}{0\%} & 0\% \\ \cline{2-8} 
\multicolumn{1}{|c|}{} & multi\_booth & \multicolumn{1}{r|}{0\%} & \multicolumn{1}{r|}{0\%} & 0\% & \multicolumn{1}{r|}{0\%} & \multicolumn{1}{r|}{0\%} & 50\% \\ \cline{2-8} 
\multicolumn{1}{|c|}{} & multi\_pipe\_4bit & \multicolumn{1}{r|}{0\%} & \multicolumn{1}{r|}{0\%} & 100\% & \multicolumn{1}{r|}{0\%} & \multicolumn{1}{r|}{100\%} & 50\% \\ \cline{2-8} 
\multicolumn{1}{|c|}{} & multipipe\_8bit & \multicolumn{1}{r|}{0\%} & \multicolumn{1}{r|}{0\%} & 0\% & \multicolumn{1}{r|}{0\%} & \multicolumn{1}{r|}{0\%} & 100\% \\ \hline
\multicolumn{1}{|c|}{\multirow{12}{*}{Advanced}} & 1x2nocpe & \multicolumn{1}{r|}{60\%} & \multicolumn{1}{r|}{40\%} & 34\% & \multicolumn{1}{r|}{100\%} & \multicolumn{1}{r|}{100\%} & 100\% \\ \cline{2-8} 
\multicolumn{1}{|c|}{} & 1x4systolic & \multicolumn{1}{r|}{40\%} & \multicolumn{1}{r|}{60\%} & 0\% & \multicolumn{1}{r|}{60\%} & \multicolumn{1}{r|}{20\%} & 100\% \\ \cline{2-8} 
\multicolumn{1}{|c|}{} & 2x2systolic & \multicolumn{1}{r|}{0\%} & \multicolumn{1}{r|}{0\%} & 0\% & \multicolumn{1}{r|}{20\%} & \multicolumn{1}{r|}{0\%} & 100\% \\ \cline{2-8} 
\multicolumn{1}{|c|}{} & 3stagepipe & \multicolumn{1}{r|}{0\%} & \multicolumn{1}{r|}{0\%} & 0\% & \multicolumn{1}{r|}{0\%} & \multicolumn{1}{r|}{0\%} & 0\% \\ \cline{2-8} 
\multicolumn{1}{|c|}{} & 4x4spatialacc & \multicolumn{1}{r|}{0\%} & \multicolumn{1}{r|}{0\%} & 0\% & \multicolumn{1}{r|}{0\%} & \multicolumn{1}{r|}{0\%} & 0\% \\ \cline{2-8} 
\multicolumn{1}{|c|}{} & 5stagepipe & \multicolumn{1}{r|}{0\%} & \multicolumn{1}{r|}{0\%} & 0\% & \multicolumn{1}{r|}{0\%} & \multicolumn{1}{r|}{0\%} & 0\% \\ \cline{2-8} 
\multicolumn{1}{|c|}{} & fsm & \multicolumn{1}{r|}{20\%} & \multicolumn{1}{r|}{0\%} & 0\% & \multicolumn{1}{r|}{0\%} & \multicolumn{1}{r|}{100\%} & 60\% \\ \cline{2-8} 
\multicolumn{1}{|c|}{} & macpe & \multicolumn{1}{r|}{0\%} & \multicolumn{1}{r|}{80\%} & 100\% & \multicolumn{1}{r|}{100\%} & \multicolumn{1}{r|}{100\%} & 100\% \\ \cline{2-8} 
\multicolumn{1}{|c|}{} & statemachine & \multicolumn{1}{r|}{0\%} & \multicolumn{1}{r|}{20\%} & 0\% & \multicolumn{1}{r|}{60\%} & \multicolumn{1}{r|}{100\%} & 100\% \\ \cline{2-8} 
\multicolumn{1}{|c|}{} & 5state\_fsm & \multicolumn{1}{r|}{20\%} & \multicolumn{1}{r|}{0\%} & 0\% & \multicolumn{1}{r|}{100\%} & \multicolumn{1}{r|}{100\%} & 60\% \\ \cline{2-8} 
\multicolumn{1}{|c|}{} & 3state\_fsm & \multicolumn{1}{r|}{0\%} & \multicolumn{1}{r|}{40\%} & 0\% & \multicolumn{1}{r|}{20\%} & \multicolumn{1}{r|}{100\%} & 100\% \\ \cline{2-8} 
\multicolumn{1}{|c|}{} & 4state\_fsm & \multicolumn{1}{r|}{60\%} & \multicolumn{1}{r|}{0\%} & 100\% & \multicolumn{1}{r|}{40\%} & \multicolumn{1}{r|}{100\%} & 80\% \\ \hline
\multicolumn{2}{|l|}{Success Rate} & \multicolumn{1}{r|}{9.38\%} & \multicolumn{1}{r|}{16.25\%} & 25.88\% & \multicolumn{1}{r|}{40.63\%} & \multicolumn{1}{r|}{59.38\%} & 71.81\% \\ \hline
\end{tabular}%
}
\end{table}
% \usepackage{multirow}
% \usepackage{graphicx}
\begin{table}[]
\caption{The next token predicts success rate.}
\label{tab:nexttoken}
\resizebox{\linewidth}{!}{%
\begin{tabular}{|l|ll|ll|}
\hline
\multicolumn{1}{|c|}{\multirow{2}{*}{\textbf{Benchmark}}} & \multicolumn{2}{c|}{GPT4-V Series} & \multicolumn{2}{c|}{Llava Series} \\ \cline{2-5} 
\multicolumn{1}{|c|}{} & \multicolumn{1}{l|}{Co-Design} & NL-Only & \multicolumn{1}{l|}{Co-Design} & NL-Only \\ \hline
1x2nocpe & \multicolumn{1}{l|}{100\%} & 100\% & \multicolumn{1}{l|}{67\%} & 100\% \\ \hline
1x4systolic & \multicolumn{1}{l|}{100\%} & 100\% & \multicolumn{1}{l|}{67\%} & 33\% \\ \hline
2state\_fsm & \multicolumn{1}{l|}{100\%} & 67\% & \multicolumn{1}{l|}{67\%} & 33\% \\ \hline
2x2systolic & \multicolumn{1}{l|}{100\%} & 100\% & \multicolumn{1}{l|}{67\%} & 67\% \\ \hline
3stagepipe & \multicolumn{1}{l|}{67\%} & 67\% & \multicolumn{1}{l|}{0\%} & 33\% \\ \hline
3state\_fsm & \multicolumn{1}{l|}{0\%} & 100\% & \multicolumn{1}{l|}{33\%} & 0\% \\ \hline
4state\_fsm & \multicolumn{1}{l|}{67\%} & 33\% & \multicolumn{1}{l|}{0\%} & 33\% \\ \hline
4x4spatialacc & \multicolumn{1}{l|}{67\%} & 67\% & \multicolumn{1}{l|}{33\%} & 0\% \\ \hline
5stagepipe & \multicolumn{1}{l|}{33\%} & 33\% & \multicolumn{1}{l|}{0\%} & 0\% \\ \hline
5state\_fsm & \multicolumn{1}{l|}{100\%} & 100\% & \multicolumn{1}{l|}{0\%} & 0\% \\ \hline
fsm & \multicolumn{1}{l|}{0\%} & 0\% & \multicolumn{1}{l|}{0\%} & 0\% \\ \hline
macpe & \multicolumn{1}{l|}{100\%} & 67\% & \multicolumn{1}{l|}{67\%} & 33\% \\ \hline
accu & \multicolumn{1}{l|}{100\%} & 100\% & \multicolumn{1}{l|}{33\%} & 33\% \\ \hline
adder\_16bit & \multicolumn{1}{l|}{100\%} & 33\% & \multicolumn{1}{l|}{0\%} & 33\% \\ \hline
adder\_16bit\_csa & \multicolumn{1}{l|}{100\%} & 100\% & \multicolumn{1}{l|}{0\%} & 0\% \\ \hline
adder\_32bit & \multicolumn{1}{l|}{67\%} & 67\% & \multicolumn{1}{l|}{0\%} & 0\% \\ \hline
adder\_64bit & \multicolumn{1}{l|}{100\%} & 33\% & \multicolumn{1}{l|}{0\%} & 0\% \\ \hline
adder\_8bit & \multicolumn{1}{l|}{100\%} & 100\% & \multicolumn{1}{l|}{67\%} & 0\% \\ \hline
div\_16bit & \multicolumn{1}{l|}{33\%} & 33\% & \multicolumn{1}{l|}{0\%} & 33\% \\ \hline
multi\_16bit & \multicolumn{1}{l|}{0\%} & 33\% & \multicolumn{1}{l|}{0\%} & 0\% \\ \hline
multi\_booth & \multicolumn{1}{l|}{100\%} & 67\% & \multicolumn{1}{l|}{67\%} & 33\% \\ \hline
multi\_pipe\_4bit & \multicolumn{1}{l|}{0\%} & 0\% & \multicolumn{1}{l|}{0\%} & 33\% \\ \hline
multi\_pipe\_8bit & \multicolumn{1}{l|}{67\%} & 67\% & \multicolumn{1}{l|}{33\%} & 0\% \\ \hline
alu & \multicolumn{1}{l|}{100\%} & 0\% & \multicolumn{1}{l|}{0\%} & 0\% \\ \hline
edge\_detect & \multicolumn{1}{l|}{67\%} & 67\% & \multicolumn{1}{l|}{33\%} & 0\% \\ \hline
freq\_div & \multicolumn{1}{l|}{100\%} & 100\% & \multicolumn{1}{l|}{33\%} & 33\% \\ \hline
Johnson\_Counter & \multicolumn{1}{l|}{67\%} & 0\% & \multicolumn{1}{l|}{33\%} & 33\% \\ \hline
mux & \multicolumn{1}{l|}{100\%} & 100\% & \multicolumn{1}{l|}{33\%} & 33\% \\ \hline
parallel2serial & \multicolumn{1}{l|}{100\%} & 100\% & \multicolumn{1}{l|}{100\%} & 33\% \\ \hline
pulse\_detect & \multicolumn{1}{l|}{100\%} & 100\% & \multicolumn{1}{l|}{0\%} & 0\% \\ \hline
right\_shifter & \multicolumn{1}{l|}{67\%} & 67\% & \multicolumn{1}{l|}{33\%} & 0\% \\ \hline
serial2parallel & \multicolumn{1}{l|}{33\%} & 67\% & \multicolumn{1}{l|}{33\%} & 0\% \\ \hline
width\_8to16 & \multicolumn{1}{l|}{33\%} & 33\% & \multicolumn{1}{l|}{33\%} & 33\% \\ \hline
Average & \multicolumn{1}{l|}{71.72\%} & 63.64\% & \multicolumn{1}{l|}{28.28\%} & 20.20\% \\ \hline
\end{tabular}%
}
\end{table}
\subsection{Ablation Study} To further reveal the difference between the natural-language model-based hardware generation and multi-modal model-based hardware generation, we compared the image-only mode, text-only mode, and mix input mode. As shown in Tab. \ref{tab:evalgeneration}, the results show that from the success rate perspective, the mix input mode is the best and the image-only mode is the worst. Specifically, the average success rate of image-only mode is 33.90\%, and the average success rate of mix input mode is 71.81\% in GPT4-V. Therefore, we recommend LLM for RTL generation with multi-modal model rather than the vision-only mode.

\section{Conclusion}

Our research underscores the significant potential of multi-modal large language models in Verilog generation. By integrating visual representations with natural language processing, we have achieved notable improvements in the generation of complex hardware designs. In addition, we propose a novel query language framework that enhances code quality and efficiency, and the comprehensive benchmark we established demonstrates a substantial increase in model performance. This approach not only advances hardware design methodologies but also provides a possible way for future research in generative AI applications within this field, marking a significant step towards more intuitive and efficient hardware design processes.
\begin{acks}
This work was supported by the National Natural Science Foundation of China under NSFC.62090024, 62222411, 62025404, 92373206, 62202453
\end{acks}
\bibliographystyle{acm}
\bibliography{reference}

\end{document}